\documentclass[twocolumn,numberedappendix,iop]{openjournal}

\usepackage[T1]{fontenc}

\usepackage[bitstream-charter]{mathdesign}
\usepackage[svgnames]{xcolor}
\usepackage[colorlinks,linkcolor=blue,citecolor=blue,urlcolor=blue]{hyperref}

\usepackage{orcidlink}
\usepackage{graphicx}

\usepackage{amsmath, amssymb}

\begin{document}

\title{Which gravitational lensing degeneracies are broken in wave-optics?}

\author{Ashish Kumar Meena\orcidlink{0000-0002-7876-4321}$^{\star}$}
\thanks{$^\star$\href{mailto:ashishmeena766@gmail.com}{ashishmeena766@gmail.com}}
\affiliation{Physics Department, Ben-Gurion University of the Negev, PO Box 653, Be’er-Sheva 84105, Israel}

\begin{abstract}

This paper studies gravitational lensing degeneracies in the wave-optics regime, focusing on lensed gravitational waves (GWs). Considering lensing degeneracies as re-scaling (or transformations) of arrival time delay surface, we can divide them into local and global types. Local degeneracies only affect the time delay surface in localized regions, whereas global degeneracies re-scale the whole time delay surface by a constant while keeping the various observed image properties unchanged. We show that local degeneracies can be broken in the wave-optics regime since lensing effects become sensitive to the overall arrival time delay surface and not only to the time delay values at the image positions. On the other hand, global degeneracies (such as similarity transformation, prismatic transformation, and mass-sheet degeneracy) multiply the amplification factor by a constant factor~(let us say, $\lambda$). However, in GW lensing, as the GW signal amplitude depends on the source distance, it turns out that~$\lambda$ is completely degenerate with the Hubble constant, similar to what we see in geometric optics. Hence, with the lensing of GWs, global degeneracies are as hard to break in wave optics as they are in geometric optics.

\end{abstract}

\maketitle

\section{Introduction}
\label{sec:intro}

With the continuous discovery of gravitational wave~(GW) signals by the LIGO-Virgo-KAGRA~(LVK) network~\citep[e.g.,][]{2024PhRvD.109b2001A}, the next highly anticipated discovery is a lensed GW event. Similar to electromagnetic~(EM) waves, GWs are also subjected to gravitational lensing~\citep[e.g.,][]{1971NCimB...6..225L, 1974IJTP....9..425O}. Multiple searches for lensed GW signals have been carried out using the existing data, but none of them has found convincing evidence for the detection of a lensed GW signal~\citep[e.g.,][]{2021ApJ...923...14A, 2024ApJ...970..191A}. Gravitational lensing of a signal with frequency~$f$ by masses~$M\lesssim10^5{\rm M_\odot}(f/{\rm Hz})^{-1}$ cannot be studied under geometric optics approximation~\citep[e.g.,][]{1986PhRvD..34.1708D, 1992grle.book.....S, 2017ApJ...835..103T} which is our go-to when studying lensing of EM waves. This implies that while studying lensing of GWs by lens masses~$\lesssim10^4{\rm M_\odot}$ in the LVK frequency range~$[10, 10^4]$Hz, wave-optical effects need to be taken into account. Similarly, wave effects are important for LISA if lens masses are~$\lesssim10^9{\rm M_\odot}$~\citep[e.g.,][]{2011MNRAS.415.2773S, 2020MNRAS.492.1127M}.

Studying GW lensing in the wave-optics regime gives us access to additional information compared to the lensing of EM waves in the geometric optics regime. First, lensing effects~\citep[such as amplification factor;][]{2003ApJ...595.1039T} become sensitive to the frequency of the GW signal. Second, we observe the amplitude of the GW signal, meaning that we also have access to the phase information, which is lost in the lensing of EM waves where we observe flux~(=|amplitude|$^2$). That said, wave effects may also show for coherent EM point sources such as fast radio bursts or pulsars~\citep[e.g.,][]{2020MNRAS.497.4956J} or in plasma lensing~\citep[e.g.,][]{2023MNRAS.525.2107J}. Owing to these additional pieces of information at our disposal, we can ask: what more can we learn about the lens while studying lensing in the wave-optics regime? Can it especially help us break various lensing degeneracies originally encountered in the lensing of EM waves in geometric optics regimes? Degeneracies in gravitational lensing refer to certain transformations that leave various observed image properties unchanged~(\citealt{1985ApJ...289L...1F, 1988ApJ...327..693G}; also see Figure~4 in~\citealt{2018A&A...620A..86W}). These transformations can be thought of as re-scaling (or transformations) of the arrival time delay surface and can be divided into global and local types~\citep{2000AJ....120.1654S}. Global degeneracies re-scale the arrival time delay surface by a constant factor such as similarity, prismatic, and mass-sheet transformations~\citep[e.g.,][]{2018A&A...620A..86W}. On the other hand, local degeneracies affect the arrival time surface in localized regions such that all lensing observables (in geometric optics regime) are invariant under these transformations, such as monopole degeneracy~\citep[e.g.,][]{2008MNRAS.389..415L, 2012MNRAS.425.1772L}. In this work, we look at the effects of lensing degeneracies on the lensed GW signals to determine which one of these can be broken under wave-optics approximation.

This work is organized as follows. Section~\ref{sec:gl_basic} reviews the gravitational lensing basics in the wave-optics regime. Section~\ref{sec:Ff} discusses the numerical method that we employ to calculate the amplification factor. In Section~\ref{sec:wo_deg}, we study various degeneracies in the wave-optics regime. We conclude and summarize our work in Section~\ref{sec:conclusions}.

\section{Lensing basics in wave-optics regime}
\label{sec:gl_basic}

The study of lensing of GWs is essentially solving the wave propagation in a curved background. In weak field limit, the background can be described by the perturbed FRW metric, given as
\begin{equation}
    ds^2 = -\left(1+\frac{2U}{c^2}\right)c^2dt^2 + a^2 \left(1-\frac{2U}{c^2}\right) d\pmb{r}^2,
\end{equation}
where~$a$ is the scale factor and~$U(\ll1)$ denotes the lensing potential of the lens. GWs upon this background are described as tensor perturbation given by~$h_{\mu\nu}$. Following the Eikonal approximation,~$h_{\mu\nu}=\phi e_{\mu\nu}$; where~$\phi$ represents the amplitude and $e_{\mu\nu}$ is the polarisation tensor of the GW. Assuming that the change in polarisation tensor is negligible due to lensing~\citep{2003ApJ...595.1039T}, the wave propagation equation for the scalar amplitude~$\phi$ in the frequency domain (up to leading order) is,
\begin{equation}
    (c^2\nabla^2 + \omega^2) \tilde{\phi} = 4\omega^2 c^{-2} U \tilde{\phi},
\end{equation}
where~$\omega=2\pi f$ and~$f$ is the frequency of the GW. The above equation can be solved using Kirchhoff diffraction integral~\citep{1999PhRvD..59h3001B}. Following~\citet{2003ApJ...595.1039T}, the amplification factor,~$F(f)$, can be defined as the ratio of lensed and unlensed~($U=0$) GW amplitudes such that in the absence of lens~$F(f)=1$. With that, under thin-lens approximation, the amplification factor can be written as,
\begin{equation}
    F(f,\pmb{y}) = \frac{1+z_d}{c} \frac{D_s}{D_d D_{ds}} \xi_0^2 \frac{f}{i} \int d^2\pmb{x} \exp[2\pi i f t_d(\pmb{x}, \pmb{y})],
    \label{eq:amp}
\end{equation}
where~$\pmb{x}=\pmb{\xi}/\xi_0$ and~$\pmb{y}=\pmb{\eta}D_d/\xi_0D_s$ are dimensionless vectors in the image and source planes, respectively. $D_d, D_{ds}, D_s$ are the angular diameter distances from observer to lens, lens to source, and observer to source, respectively. $\xi_0$ is an arbitrary length scale in the image plane, and~$z_d$ is the lens redshift. $t_d(\pmb{x}, \pmb{y})$ represents the arrival time delay surface which is given as,
\begin{equation}
    t_d(\pmb{x}, \pmb{y}) = \frac{1+z_d}{c} \frac{D_s}{D_d D_{ds}} \xi_0^2 \left[ \frac{(\pmb{x} - \pmb{y})^2}{2} - \psi(\pmb{x}) \right],
    \label{eq:td}
\end{equation}
where symbols have their usual meaning. The phase of the amplification factor in Equation~\eqref{eq:amp} is defined as $\theta_F\equiv-i\ln[F/|F|]$.

Under geometric optics approximation ($ft_d\gg1$), the integral in Equation~\eqref{eq:amp} becomes highly oscillatory, and only its stationary points contribute to the amplification factor. In this limit, using stationary phase approximation, the amplification factor can be written as,
\begin{equation}
    F(f, \pmb{y})|_{\rm geo} = \sum_j \sqrt{|\mu_j|} \exp \left( 2\pi i f t_d(\pmb{x}_j, \pmb{y}) - i \pi n_j \right),
    \label{eq:amp_geo}
\end{equation}
where~$t_d(\pmb{x}_j, \pmb{y})$ is the value of time delay for $j$-th image and~$\mu_j$ is the corresponding well-known magnification factor. $n_j$ is the Morse index having a value of 0, 1/2, and 1 for minima, saddle, and maxima images, respectively. Often, it is useful to write the time delay and frequency in dimensionless quantities, which can be defined as,
\begin{equation}
        T_s  \equiv \frac{1+z_d}{c} \frac{D_s}{D_d D_{ds}} \xi_0^2;\quad
        \tau_d(\pmb{x}, \pmb{y}) \equiv \frac{t_d(\pmb{x}, \pmb{y})}{T_s}; \quad
        \nu \equiv T_s f.
    \label{eq:dimless}
\end{equation}
With the above, the amplification factor given in Equation~\eqref{eq:amp} and~\eqref{eq:amp_geo} can be written as,
\begin{equation}
    \begin{split}
        F(\nu,\pmb{y}) &= \frac{\nu}{i} \int d^2\pmb{x} \exp[2\pi i \nu \tau_d(\pmb{x}, \pmb{y})], \\
        F(\nu, \pmb{y})|_{\rm geo} &= \sum_j \sqrt{|\mu_j|} \exp \left( 2\pi i \nu \tau_d(\pmb{x}_j, \pmb{y}) - i \pi n_j \right).
    \end{split}
    \label{qe:amp_dimless}
\end{equation}

\begin{figure*}
    \centering
    \includegraphics[width=14cm, height=12cm]{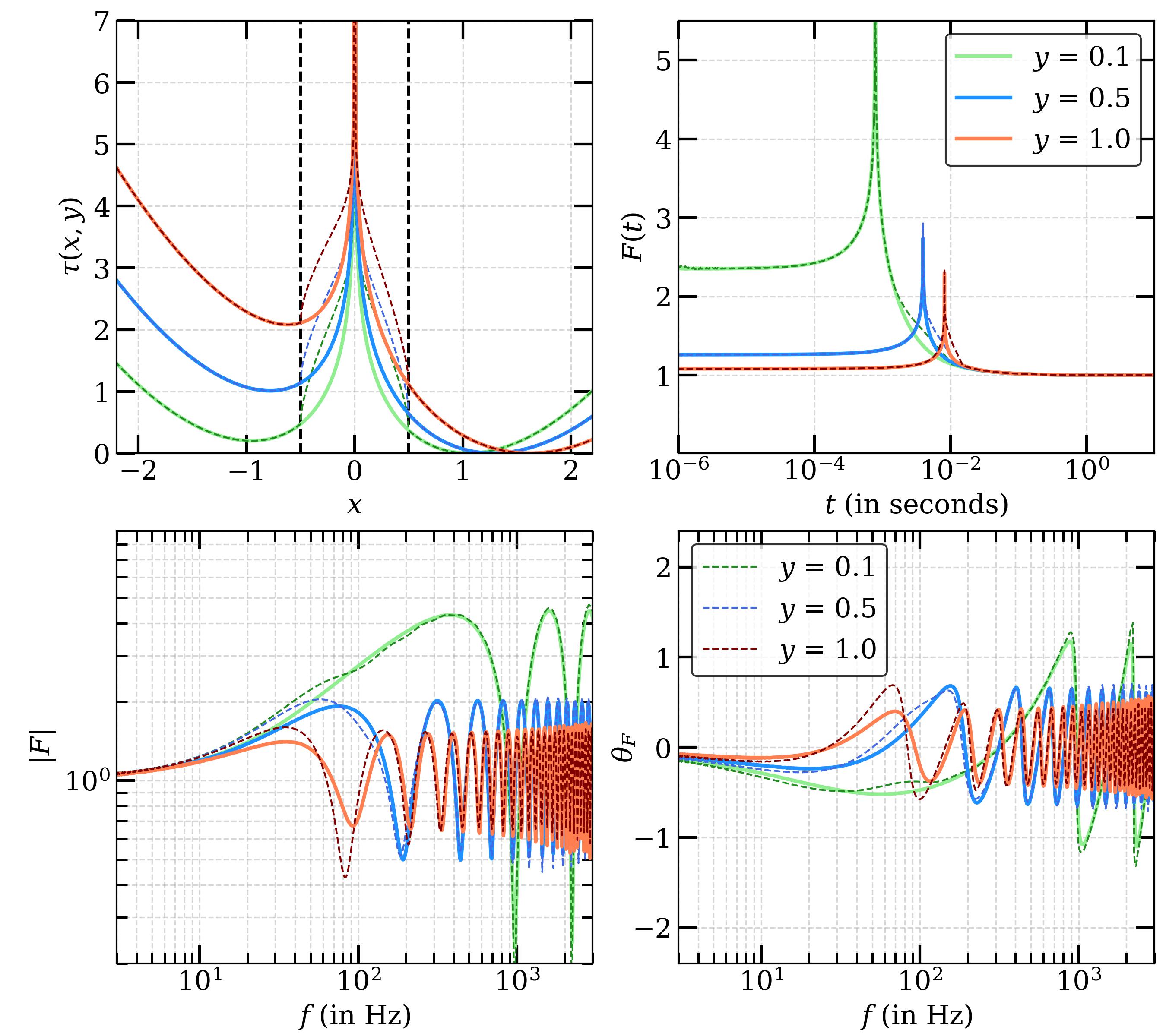}
    \caption{Comparison of various quantities for point mass lens and PM+Sphere lens with $(W, R)=(0.5, 0.5)$. In each panel, solid curves are corresponding to the point mass lens, whereas dashed curves are for the PM+Sphere lens. \textit{Top-left}. Dimensionless time delay function,~$\tau(x,y)$ for different source positions~$(y)$. The vertical dashed black lines represent the extent of the uniform-density sphere. \textit{Top-right}. (Inverse) Fourier transform of amplification factor, $F(t)$.  \textit{Bottom-left}. The absolute value of the amplification factor~$|F|$. \textit{Bottom-right}. Phase factor of the amplification factor~$\theta_F$.}
    \label{fig:monopole}
\end{figure*}

\section{Solving~$F(\lowercase{f})$}
\label{sec:Ff}

The integral in amplification factor, Equation~\eqref{eq:amp}, can only be solved analytically for very simple lens models. For example, in the case of a point mass lens~($\psi(x) = \ln|x|$), the solution can be written as~\citep{2003ApJ...595.1039T},
\begin{equation}
    \begin{split}
    F(\omega,y) = \exp \left\{ \frac{\pi \omega}{4} + \frac{i\omega}{2} \left[ \ln \left(\frac{\omega}{2}\right) - 2 \phi_{\rm m}(y) \right] \right\} \\
    \Gamma\left( 1-\frac{i\omega}{2} \right) {}_1F_1 \left( \frac{i\omega}{2}, 1; \frac{i\omega y^2}{2} \right),         
    \end{split}
    \label{eq:pm_analytic}
\end{equation}
where~$\omega\equiv2\pi\nu$ and $\phi_{\rm m}(y)=(x_{\rm m}-y)^2/2-\ln(x_{\rm m})$ with $x_{\rm m} = (y+\sqrt{y^2+4})/2$ represents the dimensionless time delay corresponding to the global minima image with compared to unlensed case. Here, the length scale~$\xi_0$ is chosen to be equal to the corresponding Einstein radius,
\begin{equation}
    \xi_E = \sqrt{\frac{4\rm{G}M}{c^2}\frac{D_d D_{ds}}{D_s}},
\end{equation}
where~$M$ is the mass of the point mass lens. Similarly, for the SIS lens, the amplification factor can be solved analytically as described in~\citet{2006JCAP...01..023M}.

For more complex lens models, Equation~\eqref{eq:amp} can only be solved numerically. One such method includes contour integration and Fourier transformation as discussed in~\citet{1995ApJ...442...67U} and used in current work. The inverse Fourier transformation of~$iF(\nu)/\nu$ can be written as,
\begin{equation}
    \begin{split}
        F(\tau') &\equiv \int_{-\infty}^{+\infty} d\nu \: \exp(-i2\pi\nu\tau') \frac{iF(\nu)}{\nu} \\
        \Rightarrow F(\tau') &= \int d^2\pmb{x} \: \delta[\tau(\pmb{x}) - \tau'],
    \end{split}
    \label{eq:Ft}
\end{equation}
where $\tau'$ represents the dimensionless time delay value. Above, in the second equation, we substituted~$F(\nu)$ from Equation~\eqref{qe:amp_dimless}. And~$F(f)$ can be written as the Fourier transformation of~$F(t)$,
\begin{equation}
    F(f) = \frac{f}{i} \int dt_d \exp(2\pi i f t_d) \: F(t_d),
    \label{eq:Ff_Ft}
\end{equation}
where~$t_d$ represents the time-delay value with respect to an arbitrary reference time, which is generally chosen such that~$t_d=0$ for global minima. Hence, the first step is to calculate the~$F(t)$ values and then take its Fourier transformation to determine the amplification factor, $F(f)$. 

As discussed in~\citet{1995ApJ...442...67U}, Equation~\eqref{eq:Ft} can be evaluated as a contour integral. The area between contours  of~$\tau_d(\pmb{x})=\tau'$ and $\tau_d(\pmb{x})=\tau'+d\tau'$ is~$A=F(\tau')d\tau'$. Also, the area between these two contours can be given as~$A=\oint{ds\:dl}$, where~$ds$ is the infinitesimal length along the contour and~$dl=d\tau'/|\nabla_{\pmb x} \tau_d|$ is the orthogonal distance between two contours at~$\pmb{x}$. Thus, by comparison,
\begin{equation}
    F(\tau') = \sum_k  \oint_{C_k}{ \frac{ds}{|\nabla_{\pmb x} \tau_d|} },
\end{equation}
where the summation is over all contours of~$\tau'$. Also, we can see that $F(\tau')$ is a smooth function except for critical values of~$\tau_d(\pmb{x})$, i.e., time delay values corresponding to lensed images in geometric optics. We refer readers to~\citet{1995ApJ...442...67U} on how to deal with these singularities. Another point is that we can only calculate~$F(t)$ for a finite range of values whereas the integral in Equation~\eqref{eq:Ff_Ft} runs in range~$[-\infty, +\infty]$. To deal with this, we use a cosine apodization function as discussed in~\citet{2019A&A...627A.130D} and \citet{2021MNRAS.508.4869M}. The above method is validated and used multiple previous works~\citep{2021MNRAS.508.4869M, 2022MNRAS.517..872M, 2024MNRAS.532.3568M}. Other methods to solve the above integral are discussed in~\citep[e.g.,][]{2019A&A...627A.130D, 2024arXiv240904606V}.

\section{Lensing Degeneracies \lowercase{vs.} wave-optics}
\label{sec:wo_deg}

As discussed in~\citet{2000AJ....120.1654S}, lensing degeneracies can be understood as the transformation of the arrival time delay surface,~$t_d(\pmb{x}, \pmb{y})$ given by Equation~\eqref{eq:td} such that the various observed image properties remain unchanged. These transformations can be broadly divided into global and local transformations. Global transformation scales the arrival time delay as a whole, whereas local transformation only modifies a part of it. Examples of global degeneracies are similarity transformation, prismatic transformation and mass-sheet degeneracy~\citep[MSD; e.g.,][]{1988ApJ...327..693G}. Similarity transformation only scales the arrival time delay surface by a constant factor~(let us say,~$\lambda$), leaving image positions~($\pmb{x}_i$) and magnifications~($\mu_i$) unchanged, i.e.,~$(t_{d,\lambda}, \pmb{x}_{i, \lambda}, \mu_{i,\lambda}) = (\lambda t_d, \pmb{x}_i, \mu_i)$. Prismatic transformation leaves all of the above three observables unchanged, i.e.,~$(t_{d,\lambda}, \pmb{x}_{i, \lambda}, \mu_{i,\lambda}) = (t_d, \pmb{x}_i, \mu_i)$. MSD, in addition to time delay, also scales the magnification factors such that their ratios are still the same, i.e.,~$(t_{d,\lambda}, \pmb{x}_{i, \lambda}, \mu_{i,\lambda}) = (\lambda t_d, \pmb{x}_i, \mu_i/\lambda^2)$. Examples of local degeneracies include monopole degeneracy and generalized mass-sheet degeneracy~(gMSD)\footnote{Here, we call gMSD a local degeneracy since we need to locally modify the time delay surface by adding monopoles. But it also does an overall rescaling of surface density similar to MSD.}, or any other transformation concocted to modify the arrival time delay surface such that they leave the observable image properties unchanged. In this section, we look at three degeneracies, namely, monopole, MSD, and gMSD, in the wave-optics regime to determine which one of these are broken.
\subsection{Monopole degeneracy}
\label{ssec:monopole}

The monopole degeneracy was first introduced in~\citet{2000AJ....120.1654S} and was further discussed in detail in~\citet{2008MNRAS.389..415L, 2012MNRAS.425.1772L}. The monopole degeneracy allows us to alter the lens mass distribution in an axis-symmetric manner in regions of the image plane which do not contain any images such that the observed image properties remain unchanged.

A simple example of monopole degeneracy can be illustrated if we compare observed image properties for a point mass lens with mass~$M$ and a point mass lens co-centred with a uniform density sphere~(PM+Sphere) whose radius is smaller than the saddle-point image position such that the total enclosed mass is again~$M$. The lensing potential for a uniform-density sphere is given as~\citep{2005PhRvD..72d3001S},
\begin{equation}
    \psi(x) = 
    \begin{cases}
    W \ln \left[ 1+\sqrt{1-\left(\frac{x}{R}\right)^2} \right] -  \\
    \qquad{} \frac{W}{3}\left[ 4-\left(\frac{x}{R}\right)^2 \right] \sqrt{1-\left(\frac{x}{R}\right)^2}, &  x \leq R \\ \\
    W \ln\left(x/R\right),              & x \geq R,
    \end{cases}
\end{equation}
where~$W$ controls the fraction of mass assigned to the uniform-density sphere and~$R$ is its radius in units of Einstein radius~$(\xi_E)$. In our work, we fix~$(W, R)=(0.5, 0.5)$, i.e., half of the total mass is redistributed as a uniform-density sphere within a radius of 0.5. 

For these two lens models, in geometric optics, all lensing observables are exactly the same, i.e.,~$(t_{d, i, \lambda}, x_{i, \lambda}, \mu_{i,\lambda}) = (t_{d, i}, x_i, \mu_i)$. This can be seen in the comparison of dimensionless time delay functions,~$\tau(x, y)$, for point mass lens and PM+Sphere lens that are shown in the top-left panel of Figure~\ref{fig:monopole}, for three different source positions. The arrival time delay surface is the same for both cases at~$x>0.5$, implying that the image positions and their magnifications will remain the same. However, the time-delay function (and contours) get modified at~$x<0.5$ since we have redistributed a certain fraction of lens mass\footnote{Redistributing all of the point lens mass as a constant density sphere will remove the central singularity and form three images. For such a case, we can distinguish point mass and PM+Sphere models even in geometric-optics regime.}. Hence, the area enclosed between contours of time delay values, $t$ and~$t+dt$ will also get modified, and its effect will be reflected in the~$F(t)$ as well as in the~$F(f)$ curves. The~$F(t)$ curves for the point mass lens and PM+Sphere lens are shown in the top-right panel of Figure~\ref{fig:monopole}, and the absolute value and phase of amplification factor are shown in the bottom row. In~$F(f)$ curves, we can clearly see the difference between the above two lens models. Hence, this implies that monopole degeneracy can be broken in the wave-optics regime since the amplification factor is sensitive to the overall shape of the time delay surface and not only to the time delay values at the image positions, as we see in geometric optics.

Here, we note that monopole~(or any other local) degeneracy can only be broken if it affects time delay contours, which affects the observed frequency range. For example, for the LVK network, the observed frequency range is~$[10, 10^4]$Hz and a local degeneracy that affects time delay values~$t_d\gg1/f$ will not affect amplification factor values in the observed frequency range.

\subsection{Mass-sheet degeneracy~(MSD)}
\label{ssec:msd}

Perhaps the most famous degeneracy in gravitational lensing is the MSD, which was originally referred to as magnification transformation in~\citet{1988ApJ...327..693G} and later also referred to as steepness degeneracy in~\citet{2006ApJ...653..936S}. MSD re-scales the existing surface density by a constant factor~$\lambda$ and adds a constant mass-sheet with density~$(1-\lambda)\Sigma_{\rm cr}$ such that the total lensing convergence can be written as,
\begin{equation}
    \kappa_\lambda (\pmb{x}) = \lambda\:\kappa(\pmb{x}) + (1-\lambda).
\end{equation}
MSD leaves the image position and magnification ratios unchanged but multiplies the individual image magnification by a factor such that~$\mu_\lambda = \mu/\lambda^2$. It can be easily shown that under MSD, the source position is re-scaled as $y_\lambda = \lambda y$. Hence, the MSD can also be written as,
\begin{equation}
    \begin{split}
        y_\lambda &= \lambda y, \\
        \psi_\lambda (\pmb{x}) &= \lambda \: \psi(\pmb{x}) + (1-\lambda)\frac{|\pmb{x}|^2}{2}.
    \end{split}
\end{equation}
With the above, the dimensionless time delay function in Equation~\eqref{eq:dimless} transforms as,
\begin{equation}
    \tau_{d,\lambda}(\pmb{x}, \pmb{y}) = \lambda \: \tau_d(\pmb{x}, \pmb{y}) - \frac{\lambda(1-\lambda)}{2} |\pmb{y}|^2.
    \label{eq:tau_lambda}
\end{equation}
Here, we see that under mass-sheet transformation, the time delay function gets scaled by the same factor~$\lambda$ along with an additive factor independent of image positions. The constant factor drops once we choose a reference point~(for example, setting the global minima to zero time), leading us to,
\begin{equation}
    \tau_{d, \lambda} = \lambda\tau_d,
\end{equation}
which is just an overall scaling of the time delay. With this, the dimensionless amplification factor can be written as,
\begin{equation}
    \begin{split}
    F_\lambda(\nu, \pmb{y}) &= \frac{\nu}{i} \int d^2\pmb{x} \exp[2\pi i \nu \lambda \tau_d(\pmb{x}, \pmb{y})] \\
    \Rightarrow F_\lambda(\nu, \pmb{y}) &= \frac{1}{\lambda} F(\lambda \nu, y).
    \end{split}
    \label{eq:msd_Ff}
\end{equation}
Hence, MSD leads to a re-scaling of dimensionless frequency,~$\nu$, and the overall amplification factor in the wave-optics regime. Although, in the above analysis, we only considered MSD, the above equation applies to other global degeneracies as they also re-scale the arrival time delay surface by a constant.

Now, the question is, can we break the MSD with the lensing of GWs in wave optics? In the frequency domain, lensed and unlensed GW waveforms are related to each other as,
\begin{equation}
    \begin{split}
    h_L(f) &= F_\lambda(f, \pmb{y}) \times h_U(f) \\
           &= \frac{1}{\lambda} F(\lambda T_s f, \pmb{y}) \times h_U(f), \\
           &= \frac{1}{\lambda} F(\lambda T_s f, \pmb{y}) \: A(f;\mathcal{M}, D_s, z_s) \: e^{i \varphi(f;\pmb{p})},
    \end{split}
    \label{eq:hl_msd}
\end{equation}
where in the last equation, we substituted the general form of GW waveform~\citep[e.g.,][]{2003ApJ...595.1039T} where~$\mathcal{M}$ denotes the chirp mass, $z_s$ is the source redshift and $\pmb{p}$ represents a source binary parameter vector controlling the phase of the GW signal. Since~$A\propto1/D_s$, we can re-scale the Hubble constant such that~$H_{0,\lambda} = H_0/\lambda$ leading us to,
\begin{equation}
    h_L(f) = F(T_s(H_{0,\lambda}), f, \pmb{y}) \: A(f; \mathcal{M}, D_s(H_{0,\lambda}), z_s) \: e^{i \varphi(f;\pmb{p})}.
\end{equation}
Although Equation~\eqref{eq:hl_msd} gives an impression that MSD can be broken since~$\lambda$ comes out as an overall scaling of~$F(f)$ or~$h_L(f)$ but since~$h_U(f)\propto1/D_s$, it remains completely degenerate with the Hubble constant similar to what we have in lensing of EM waves in geometric optics\footnote{In principle, the~$1/\lambda$ factor is also degenerate with the chirp mass~($\mathcal{M}$). But for simplicity, we only focus on the degeneracy with the Hubble constant~($H_0$).}. This suggests that even in wave optics with lensing of GWs, MSD (or any other global degeneracy) is as hard to break as it is in the geometric-optics regime, which seems to contradict what was found in~\citet{2021PhRvD.104b3503C} and~\citet{2024arXiv240803856C}.

\subsection{Generalized mass-sheet degeneracy~(gMSD)}
\label{ssec:gmsd}

As discussed in the previous sub-section, in MSD, we add a constant mass density, which is a multiple of the critical density~($\Sigma_{\rm cr}$). However, exactly due to the fact that it is multiple of critical density, it can be broken if we have two lensed sources since critical density is redshift dependent. \citet{2008MNRAS.386..307L, 2012MNRAS.425.1772L} showed that one could construct a so-called generalized mass-sheet degeneracy~(gMSD), which cannot be broken even in the presence of multiple lensed sources. This is done by first rescaling~$\kappa(\pmb{x})$ for each lensed source mass-sheet transformation and then adding monopoles to compensate for each mass-sheet such that each of the lensed sources will only be subjected to their own MSDs. This makes the gMSD a local degeneracy according to our definition as it re-scales the time delay surface in a position-dependent manner. As shown in sub-section~\ref{ssec:monopole}, adding monopoles essentially modifies the time delay surface (although not at the image position), and these changes will again be reflected in the~$F(t)$ and~$F(f)$ curves. Hence, we can again argue that, in wave-optics, the gMSD will be broken.

\section{Summary \& Conclusion}
\label{sec:conclusions}

In this work, we have studied various lensing degeneracies in the wave-optics regime, primarily focusing on lensed GW signals. We have explicitly shown that monopole degeneracy breaks in the wave-optics regime since its effects are captured in the amplification factor. Based on that, we argue that any local degeneracy can be broken in the wave-optics regime since, by definition, they will modify the arrival time delay surface~(although not at the image positions) as well as~$F(f)$. As mentioned in~\ref{ssec:monopole}, the frequency range that is used to break the monopole degeneracy is directly related to the substructure properties as it needs to affect the corresponding time delay contours.

Global degeneracies, on the other hand, are more complex. Under a global degeneracy, the arrival time delay surface gets re-scaled by a constant factor~$\lambda$, and the overall amplification factor further gets re-scaled by~$1/\lambda$. This gives the impression that global degeneracies can also be broken in GW lensing in the wave-optics regime. However, as we show in Section~\ref{ssec:msd}, it is not true since GW signal amplitude also depends on the distance to the GW source and by rescaling the Hubble constant~($H_0$), we can completely hide the global degeneracies, making them as hard to break with GW lensing in the wave-optics regime as they are under geometric-optics approximation in EM lensing.

\section*{Acknowledgements}
I am grateful to Prasenjit Saha for many useful discussions and comments. I also thank Jose Diego and Jenny Wagner for their useful comments. Author also thanks the reviewer for useful comments. I acknowledge support by grant 2020750 from the United States-Israel Binational Science Foundation (BSF), grant 2109066 from the United States National Science Foundation (NSF), and the Ministry of Science \& Technology, Israel. This research has made use of NASA’s Astrophysics Data System Bibliographic Services.

\bibliographystyle{aasjournal}
\bibliography{Reference}

\end{document}